\documentclass[fleqn,usenatbib,referee]{mnras}
\usepackage{newtxtext,newtxmath}

\usepackage[T1]{fontenc}
\usepackage{ae,aecompl}

\usepackage{graphicx}	
\usepackage{amsmath}	
\usepackage{epsfig}
\usepackage{url}
\usepackage{widetext}
\usepackage[normalem]{ulem}
\usepackage{lineno}

\title[October 2022 flare in OJ 287]{The October 2022 flare in OJ 287 and the mass of its primary black hole}
\author[M. Valtonen et al.]{Mauri J. Valtonen,$^{1,2}$\thanks{E-mail: mvaltonen2001@yahoo.com (MJV)} 
Staszek Zola,$^3$ Manpreet Singh,$^{4}$  Andrei V. Berdyugin,$^2$
\newauthor{
Kari Nilsson,$^1$ A. Gopakumar,$^5$ Alok C. Gupta,$^6$ Tapio Pursimo,$^7$  Alexandr E. Volvach,$^8$
}
\newauthor{Marek Drozdz,$^{9}$ Waldemar Ogloza,$^{9}$ Rene Hudec,$^{10,11}$ Martin Jel\'{\i}nek,$^{11}$ Jan \v{S}trobl,$^{11}$}
\newauthor{Michal Zejmo,$^{12}$ Stefano Ciprini,$^{13,14}$ Daniel E. Reichart,$^{15}$ Vladimir V. Kouprianov,$^{15}$}
\newauthor{Alberto Sadun,$^{16}$ Markus Mugrauer,$^{17}$ Katsura Matsumoto,$^{18}$ Ryo Imazawa,$^{19}$}
\newauthor{Makoto Uemura,$^{20}$ and Lankeswar Dey$^{21}$}
\\
$^1$ FINCA, University of Turku, Turku, Finland\\
$^2$ Tuorla Observatory, Department of Physics and Astronomy, University of Turku, Turku, Finland\\
$^3$ Astronomical Observatory, Jagiellonian University, ul. Orla 171, 30-244 Krakow, Poland\\
$^4$ Department of Physical Sciences, Indian Institute of Science Education and Research (IISER) Mohali, S.A.S. Nagar, Punjab, India\\
$^5$ Department of Astronomy and Astrophysics, Tata Institute of Fundamental Research, Mumbai, India\\
$^6$ Aryabhatta Research Institute of Observational Sciences (ARIES), Manora Park, Nainital 263001, India\\
$^7$ Nordic Optical Telescope, Apartado 474, E-38700 Santa Cruz de La Palma, Spain\\
$^8$ Radio Astronomy Laboratory of Crimean Astrophysical Observatory, Katsively RT-22, Crimea\\
$^{9}$ Mt. Suhora Observatory, University of the National Education Commission,
Krakow, ul. Podchorazych 2, 30-084 Krakow, Poland\\
$^{10}$ Czech Technical University, Faculty of Electrical Engineering, Prague, Czech Republic\\
$^{11}$ Astronomical Institute (ASU CAS), Ond\v{r}ejov, Czech Republic\\
$^{12}$ Kepler Institute of Astronomy, University of Zielona Gora, Lubuska 2, 65-265 Zielona Gora, Poland\\
$^{13}$ Instituto Nazionale di Fisica Nucleare (INFN) Sezione di Roma Tor Vergata, Via della Ricerca Scientifica 1,     \\00133, Roma, Italy\\
$^{14}$ ASI Space Science Data Center (SSDC), Via del Politecnico, 00133, Roma, Italy\\
$^{15}$ University of North Carolina at Chapel Hill, Chapel Hill, North Carolina, NC 27599, USA\\
$^{16}$ Department of Physics, University of Colorado, Denver, CO 80217, USA\\
$^{17}$ Astrophysikalisches Institut und Universitäts-Sternwarte, Schillergäßchen 2, D-07745 Jena, Germany\\
$^{18}$ Astronomical Institute, Osaka Kyoiku University, 4-698 Asahigaoka, Kashiwara, Osaka, 582-8582, Japan\\
$^{19}$ Department of Physics, Graduate School of Advanced Science and Engineering, Hiroshima University,\\ 1-3-1 Kagamiyama, Higashi-Hiroshima, Hiroshima 739-8526, Japan\\
$^{20}$ Hiroshima Astrophysical Science Center, Hiroshima University, 1-3-1 Kagamiyama, Higashi-Hiroshima,\\ Hiroshima 739-8526, Japan\\
$^{21}$ Institute of Astrophysics, Foundation for Research and Technology Hellas, GR-70013 Vassilika Vouton, Greece
\\}

\begin{document}
\date{Accepted XXX. Received YYY; in original form ZZZ}
\pagerange{\pageref{firstpage}--\pageref{lastpage}}
\pubyear{2026}
\maketitle
\label{firstpage}
\begin{abstract}
The bright blazar OJ~287 has demonstrated a sequence of flares, which are well 
explained by a quasi-Keplerian orbit model. The flares are associated with 
the impact of the secondary on the accretion disk of the primary. The orbit 
must precess in order to produce the correct sequence of flares, and from the 
precession rate we calculate the mass of the primary. This precession rate gives 
the mass of the primary $M_{BH} = (18.35\pm0.05) \times 10^9 M_{\odot}$. Two kinds 
of flares have been identified: direct flares from the impacts, and tidal flares 
arising from an increased accretion flow into the jet. The precession rate and 
the primary black hole mass may be independently determined from both sets of flares; 
the tidal flare of October 2022 was recommended for an intense campaign for 
this reason. This paper describes these observations over a wide spectral range. We show that the October 2022 flare fits the expectations for 
a tidal flare and thus supports the earlier determination of the mass of the 
binary black hole system in OJ 287. The mass of the primary may also be deduced 
from secondary indicators such as the correlation with the hydrogen line strength 
and the black hole mass. These studies require that the mass is 
above $M_{BH} \sim 10^{10} M_{\odot}$, but do not specify the value more exactly. 
\end{abstract}

\section{Introduction}
\label{sec:intro}
OJ~287 is a bright blazar situated at a redshift of $z = 0.306$ \citep{1985PASP...97.1158S}. Its optical observations date back to 1888, and the extended optical lightcurve spanning 137 years shows intriguing quasi-periodic variations with periods of about 12 and 55 yr years, respectively \citep{val06b}.
These unique magnitude variations in the optical light curve of OJ~287 can  be explained with the help of a binary black hole (BBH) central engine model, see \cite{2021Galax..10....1V}
and references therein  \citep{LV96,1997ApJ...484..180S,dey19a}. 

According to this model, a supermassive secondary black hole (BH) is orbiting around a much more massive primary BH in a precessing eccentric orbit with a redshifted orbital period of $\sim$12 years \citep{dey18}. 
The orbital plane is assumed to be almost perpendicular to the accretion disc of the primary BH, and the flares happen when the secondary BH collides with the disc. This model has been very successful in predicting the major flares in the optical light curve of this unique blazar since 1982 \citep{val08,val11a,val16,laine20}.

Figure 1 illustrates the view of OJ~287 based on intensive observations and modeling over more than 40 years \citep{2024ApJ...968L..17V}. The system consists of two black holes with a mass ratio of about 122, the accretion discs around each black hole, the corona of the primary, and the jets coming out of both black holes. In this study we consider the mass flow from the impact site to the center of the accretion disc and then to the jet. Previously this mass flow has been simulated and estimated as few $M_{\odot}$ per year, comparable to the average accretion rate of the primary of about $6 M_{\odot}$ per year \citep{2019ApJ...882...88V,2023MNRAS.521.6143V}. The extra mass flow may cause an increase in the jet power by a large factor after the disc crossings \citep{2025ApJ...993L..22R,2026arXiv260108080C}.

\cite{pih13}
calculate the flare structures arising from impacts in the apocenter part of the binary orbit. These complement the structures calculated for the pericenter flares previously \citep{1997ApJ...484..180S}. The three apocenter flares modeled are the 2005, 2015 and 2022 flares. The model uses a viscous disc which is perturbed by the secondary black hole, including direct impacts on the disc. The model for the 2015 impact has been discussed previously by \cite{2021Galax..10....1V}.

The model says in essence that the flare comes in different phases: a bremsstrahlung flare (also called an impact flare) from a hot bubble of gas which has been pulled out at the impact, followed by a secondary flare from the same bubble \citep{2019ApJ...882...88V}, and finally a synchrotron flare from the jet \citep{2023Galax..11...82V}, so called tidal flare. The impact flare is observed at the time when the bubble becomes optically thin \citep{LV96,1998ApJ...507..131I}. 
On the basis of Figure 2, published prior to the observing campaign \citep{2021Galax..10....1V}, 
the impact flare of 2022 was not expected to be visible from any Earth-based facility. 
Therefore, a major part of 2022 campaign of OJ~287 was aimed at the tidal flare of October 2022. 

Due to a misunderstanding \citep{2024RNAAS...8..276V}, some authors have thought that the orbit solution was recalculated after \cite{dey18} and that the new solution shifts the impact flare to October 2022, and that the campaign was about observing this impact flare \citep{Komossa_2023b}. Actually, no recalculation was done, and the misunderstanding came from a sentence in an early draft of \cite{2023MNRAS.521.6143V}. 
At the end of the paper we discuss what effect this misunderstanding, as well as other independent scientific points based on alternative mass determination methods and their own multi-wavelength observations, had on the estimated mass of the primary.

The primary mass given by \cite{dey18} is $M_{BH} = (18.35\pm0.01) \times 10^9 M_{\odot}$. The high accuracy of this solution and why the error limits may actually be somewhat wider will also be discussed in Section 5.

\begin{figure}
    \centering
    \includegraphics[angle = 0,width=0.47\textwidth]{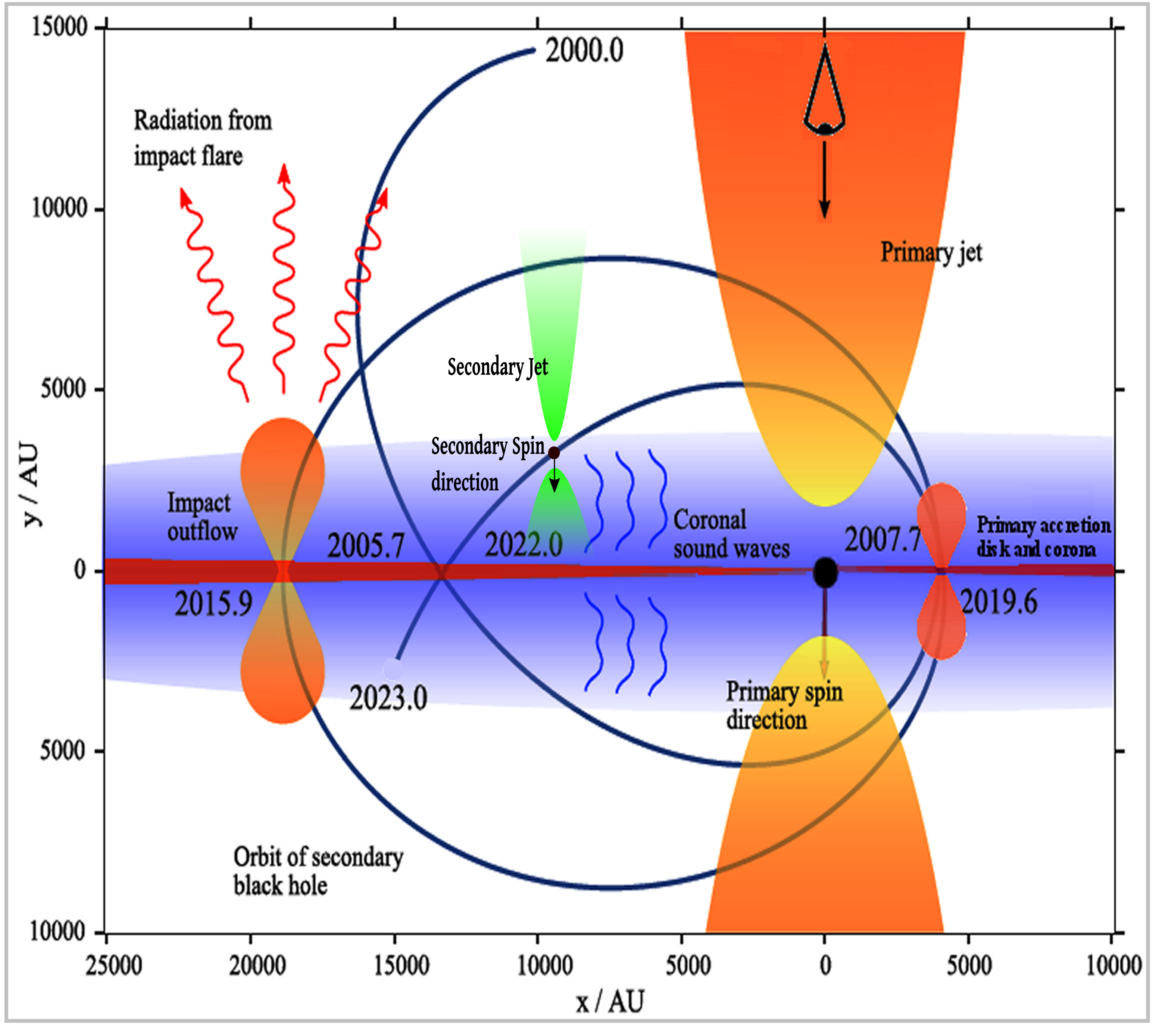}
    \caption{The binary model of OJ~287. This is an edge-on view of the accretion disc surrounding primary black hole at the coordinate origin. The thin disc is shown in red, and it is surrounded by a corona (violet). The orbit of the secondary is perpendicular to the disc, and is displayed from year 2000 to year 2023. Twin-jets arise from both black holes, the primary jets shown in yellow, and the secondary jets in green. During disc impacts bubbles of shocked gas are blown out of the disc (shown in orange), and they produce short-lived prominent flares that act as markers of the progress of the orbital motion. The system is viewed close to the rotation axis of the disc, from above. The figure has appeared previously in Valtonen et al., The Astrophysical Journal Letters, Vol. 968, p. L17-L25.}
    \label{fig:illu1}
\end{figure}

\noindent
Besides the dynamical mass from the complete orbit solution by \cite{dey18}, accurate in primary mass to better than one percent \citep{laine20}, there have been other less accurate ways to determine the masses of the two black hole components in OJ~287. Reverberation mapping, megamaser kinematics, and star and gas kinematics are the main techniques used to estimate supermassive black hole masses \citep{2004ASPC..311...69V}. Megamasers are only detectable in practically edge-on 
sources, the reverberation mapping approach requires the identification of higher-ionization emission lines 
from gas near the black hole, and the star and gas kinematics methods require high spatial resolution spectroscopy 
of the host galaxy. OJ 287's optical spectra are mostly featureless continuum and show only weak 
emission lines, making it very difficult to calculate the black hole mass using spectroscopy-based approaches. 
Additionally, as blazars are almost face-on sources, the megamaser technique is similarly irrelevant.

The black hole mass estimation techniques either use well-known empirical relationships between the black hole mass and the velocity dispersion or mass of the host galaxy's bulge, or they are approximations to the 
reverberation mapping approach that still depend on the existence of a well-measured strong emission line
\citep{2006ApJ...641..689V}. The mass of a BL Lac object's supermassive black hole can also be 
estimated using the variability timescales and the period of quasi periodic oscillations detected in 
time series flux variations \citep{2009ApJ...690..216G}.

\cite{2013MNRAS.434.3122P} discuss the variability of OJ~287 and how it may be related to the black hole mass. By studying optical variability timescales of OJ 287 on multiple occasions, the black hole mass was found in the range 
of (0.65 -- 1.67) $\times \ \rm{10}^{8} \rm{M}_{\odot}$
\citep{2012NewA...17....8G}. The value brackets the secondary black hole mass of (1.5$\pm$0.1) $\times \ \rm{10}^{8} \rm{M}_{\odot}$ from the orbit solution, accurate to better than 10 percent. In one of the clearest quasi-periodic oscillation detections the mass was estimated to be 1.46 $\times \ \rm{10}^{8} \rm{M}_{\odot}$, assuming a maximally rotating black hole \citep{2012NewA...17....8G}. This agrees perfectly with what is known about the secondary black hole in OJ~287 \citep{2024ApJ...968L..17V}.

The R-band light curve of the 2022 October flare in OJ~287 was reported in \cite{2023MNRAS.521.6143V}
without any discussion or other information about its possible origin. The purpose of this paper is to complement this information, and to see the implications for the mass of the binary black hole system. The result is compared with some other, less accurate methods of black hole mass determination.

\section{Tidal flares}

The first concrete calculation with regard to the tidal flare of 2022 October is seen in the plot of Figure 2 \citep{2021Galax..10....1V}. By accident, the 2005 and 2022 flares result from a disk impact at the same distance from the primary, and thus it is a fair assumption that the two light curves should also be similar. The tidal flare in 2005 was extended and went on until early 2006. By a time axis transformation (the 2022 time axis is plotted on top of the figure), the corresponding part of the 2022 flare was expected in October 2022. However, based on the simulation by \cite{pih13}, the 2022 flare should have been quite bit smaller than the 2005 tidal flare (exactly how much smaller will be calculated below).

\begin{figure}
    \centering
    \includegraphics[width=0.5\textwidth]{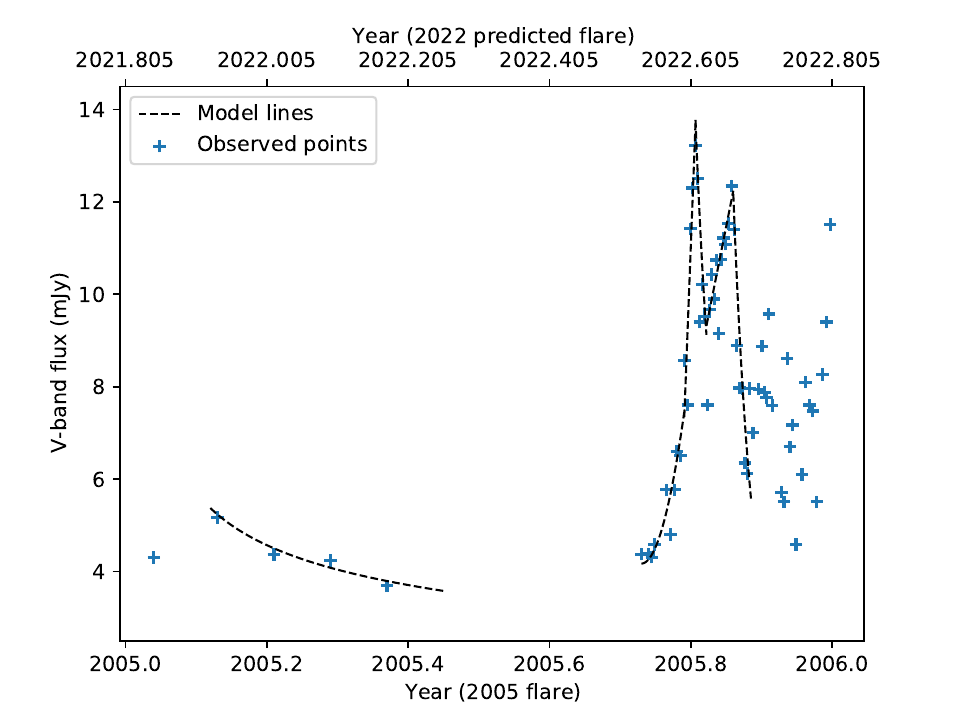}
    \caption{The V-band optical light curve of OJ~287 during  2005 in mJy units. Additionally, the curve
    provides the expected light-curve of OJ~287 during 2022; its epochs are labeled on the upper x-axis using the 
     time shift of \citet{dey18}
     . Note that OJ~287 is generally not observable during the fraction from .5 to .7 of any calendar year because OJ~287 is then too close to the Sun for optical photometry. In 2022 the tidal flare peaking at 2022.8 was the first detection opportunity.}
    \label{fig:2005_lc_fit}
\end{figure}

As to the other expected properties of the 2022 October flare, we must rely on earlier observations of similar tidal flares. There are two such flares: the 2005 tidal flare, that we just mentioned, and the 2016 tidal flare. In addition, both from previous observations and from theory, we expect a second component of the tidal flare which is even bigger than the first \citep{2017Galax...5...83V}. Thus we will report the data of the 2023 tidal flare also, and match it with its counterpart 2016/2017 tidal flare, here called the 2016-2 flare. It is not clear how to isolate the second component of the 2005/2006 tidal flare from the 2007 impact flare, and we will not study it here.
The most detailed modeling has been done for the 2016/2017 double flare. It is explained as mass flow that starts near the 2015 impact site, and arrives at the jet a little more than a year later (see Figure 3, and \cite{2021Galax..10....1V}).

\begin{figure}
\centering
\includegraphics[width=0.5\textwidth]{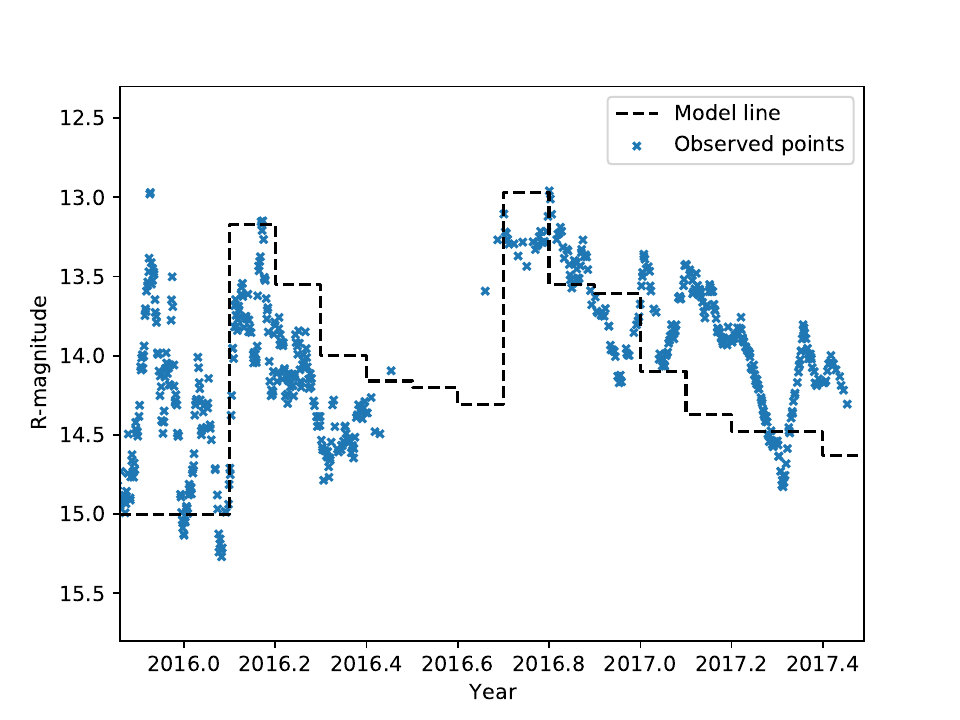}
\caption{Simulation of the response of the accretion disk to the crossing of the secondary black hole through the disk before the 2015 bremsstrahlung flare. Particles, which escape the disk are counted, and they are placed on the time axis assuming that the particle contributes to the light curve after traveling the distance of $\sim 15,000$ AU with a constant speed of $0.22 c$ where $c$ is the speed of light. The exact value of the travel distance depends on the radial distance from the center where the particle is released from the disk. The particle count in each time box is turned into a magnitude, assuming that the relation between the particle number and the total brightness in each box is linear. The first flare in observations at the end of 2015 is different (impact flare) and is not modeled by this process.\label{fig:2016-2017_lc}}
\end{figure}

Tidal flares have been quantitatively calculated in \cite{pih13}. The numbers of particles that escape from the disk per time interval are shown in Figure 4, the bottom panel, of \cite{pih13}. The scale is in mJy units, but this is arbitrary and has to be fixed by comparison with observation. The second column in Table 1 gives these numbers in mJy units. In the third column the values have been adjusted to correspond to the scale in the R-band observations, after the base level has been deducted. We have done this by using a factor of three.

\begin{figure*}
 \centering
 \includegraphics[scale=0.45,angle=270]
{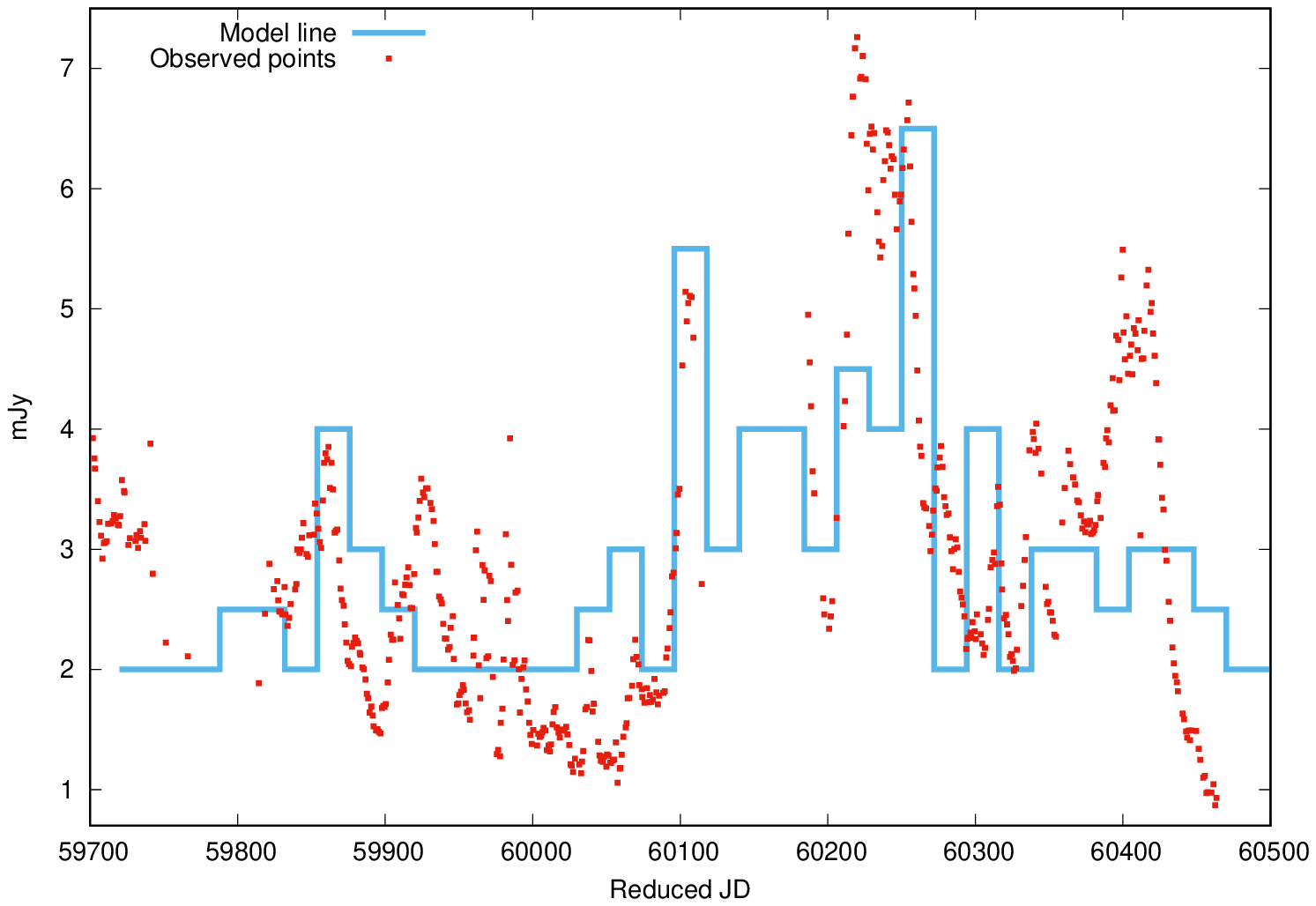} 
    \caption{R-band observations during 2022 and 2023 flares in one day bins (points) compared with the 2013 model calculation (line). In general outline, we expected two components in the light curve: the first one from August 28 to November 16, 2022, peaking at October 17, and the second one from September 12, 2023 to January 10, 2024, peaking at November 11, 2023. The observed points confirmed the first peak while the second one came a month earlier than expected. Notice that the position of the second theoretical peak is influenced by small-number statistics, and within statistical fluctuations, it agrees with observations. The light curves of the two components are shown in greater detail in Figures 5 and 6.} 
    \label{fig 1_1}
\end{figure*}

\begin{table}
\caption{Model flux and observed peak flux of 2005, 2016, and 2022 tidal flares.} 
\begin{tabular}{ c c c c c}
\hline
Flare epoch & model mJy & adjusted & observed mJy & distance au\\ 
\hline
2005  & 2.3  &   6.9  & 8.0  & 9750 \\ 
2016  & 7.9  &   19.0 & 12.0 & 15000\\
2016-2& 4.8  &   11.5 & 12.0 & 15000\\
2022  & 0.15 &   1.8  & 2.4  & 9000\\  
2023  & 0.4  &   5.0  & 6.5  & 10500\\
\hline
\multicolumn{5}{{p{0.6\textwidth}}}{\textit{Column 1: flare epoch, Column 2: the flux measured from Pihajoki et al. (2013), Column 3: adjusted flux so that the disk density remains constant with time, Column 4: observed flux, Column 5: impact distance from the centre.}}
\end{tabular}

\end{table}
However, this is not the whole story. The simulation lost particles from the disc during the runs and they were not replaced. Therefore part of the decline in numbers representing the 2022 third flare with respect to 2005 comes from this computational factor. It may be corrected by comparing the numbers in the impact flares, as they should really be the same in 2005 and 2022. Their representative numbers are in the ratio 3:1. Correcting for this, the actual size of the 2022 tidal flare should be about 1/3 of the 2005 tidal flare.

Also we need to consider a realistic radial density distribution in the disc \citep{2019ApJ...882...88V,2021Galax..10....1V}. \cite{pih13}
used a uniform surface density model with a sudden cut-off while in fact the density should be decreasing gradually with radial distance. This makes the 2016 flare too bright in the model of \cite{pih13}.

In the new simulations, reported by \cite{2021Galax..10....1V},
the theoretical 2016 tidal flare becomes smaller and is comparable to the 2016-2 flare (see Figure 3). Its adjusted value is reported for the epoch 2016-2 on the third line of Table 1, and is 11.5 mJy. It is close to the observed 12 mJy.

The number of particle escapes per 20 day bin in the \cite{pih13}
calculations is presented in Figure 4. The bins have been moved forward by about 0.5 years, corresponding to the time of travel from the impact site to the jet, using a speed which is about 20\% higher than the calculated speed for the 2016/2017 flares, $0.22 c$, where $c$ is the speed of light \citep{2021Galax..10....1V}. The observed R-band data is shown for comparison.

The travel time to the jet depends on the average speed in the corona as well as on the distance. The coronal speed should be proportional to the Kepler speed at the start of the flow towards the jet.

Note that the particles representing the 2022 flare actually have a shorter distance to travel to the jet than the particles representing the 2023 flare (see Column 5 in Table 1). This is because the secondary moves from the direction of the primary outwards at the time of impacts (Figure 1), and starts its tidal influence on the disc first closer to the center, while the later loss of mass comes at a greater distance. Taking care of the different travel times makes the separation between the two peaks somewhat longer (by about 15\%) than in the simulations since this difference is not included in the model. We have taken care of this in Figure 4 by moving the theory line forward at the 2023-peak by 66 days. Another more subtle change from the model is the increase of the box size from 20 days to 22 days to account for the motion of source counter to the particle flow. A quantitative calculation of these effects would require a new simulation like in \cite{2021Galax..10....1V}.

\section{Comparison of tidal flares}
\label{observational_data}

\subsection{Swift Observations of OJ 287 in 2016 and 2022/2023}

Despite intensive efforts along several decades of observational and theoretical research, there are still many uncertainties and open questions related to how jets are produced and what is their composition \citep{2021AN....342..727R}. From intensive monitoring of quasars it appears that the variability is associated with shocks in jets \citep{1985ApJ...298..114M,1992A&A...254...71V} where the magnetic field is compressed at the shock front. This naturally explains the increased synchrotron radiation flux as well as the higher degree of polarization at flux peaks. We expect that this will be true also for the tidal flares.

However, the real signature of tidal flares is the relation between x-ray and optical variability. There is no clear explanation why it happens, but in the best studied tidal flare of OJ~287, peaking on May 24, 2020, there was no x-ray variability that would match the prominent optical flare \citep{2022MNRAS.513.3165K}. The 2020 tidal flare is the biggest tidal flare in the theoretical light curve of \cite{sun97} for the period of 2000 - 2030. It may not be an accident that this flare is also one of the biggest in the V/X - light curve, where V is the optical V-band flux and X is the x-ray flux in Swift observations of OJ~287. Even though we do not understand the astrophysical origin of the V/X flares, we may use them to identify tidal flares.

The Swift telescope has monitored OJ~287 \textbf{intensively} since late 2015 \citep{2021MNRAS.504.5575K}. Thus, we have data for four of the tidal flare epochs in this study. In Figures 5 and 6 we plot the data for three of the bands, x-ray (called X-band), ultraviolet W2 channel, and optical B-band, as a ratio to the V-band flux. The V/X-ratio behaves similarly at both epochs, showing a clear peak at the R-band flux maximum (shown at the bottom panel). This indicates that x-rays do not participate in the flare, in any of the four cases. Generally, we would expect the V and X channels to follow each other \citep{2021Galax..10....1V}.
In contrast, the V/W2- and V/U-ratios are constant within the error bars, i.e. the optical spectral index does not change detectably at the blue end of the spectrum during the flare.

The V/X-ratio rises about by about a factor of three during tidal flares, as compared with its pre- and post-flare levels (Figure 7).  

\begin{figure*}
 \centering
 \includegraphics[scale=0.4]
{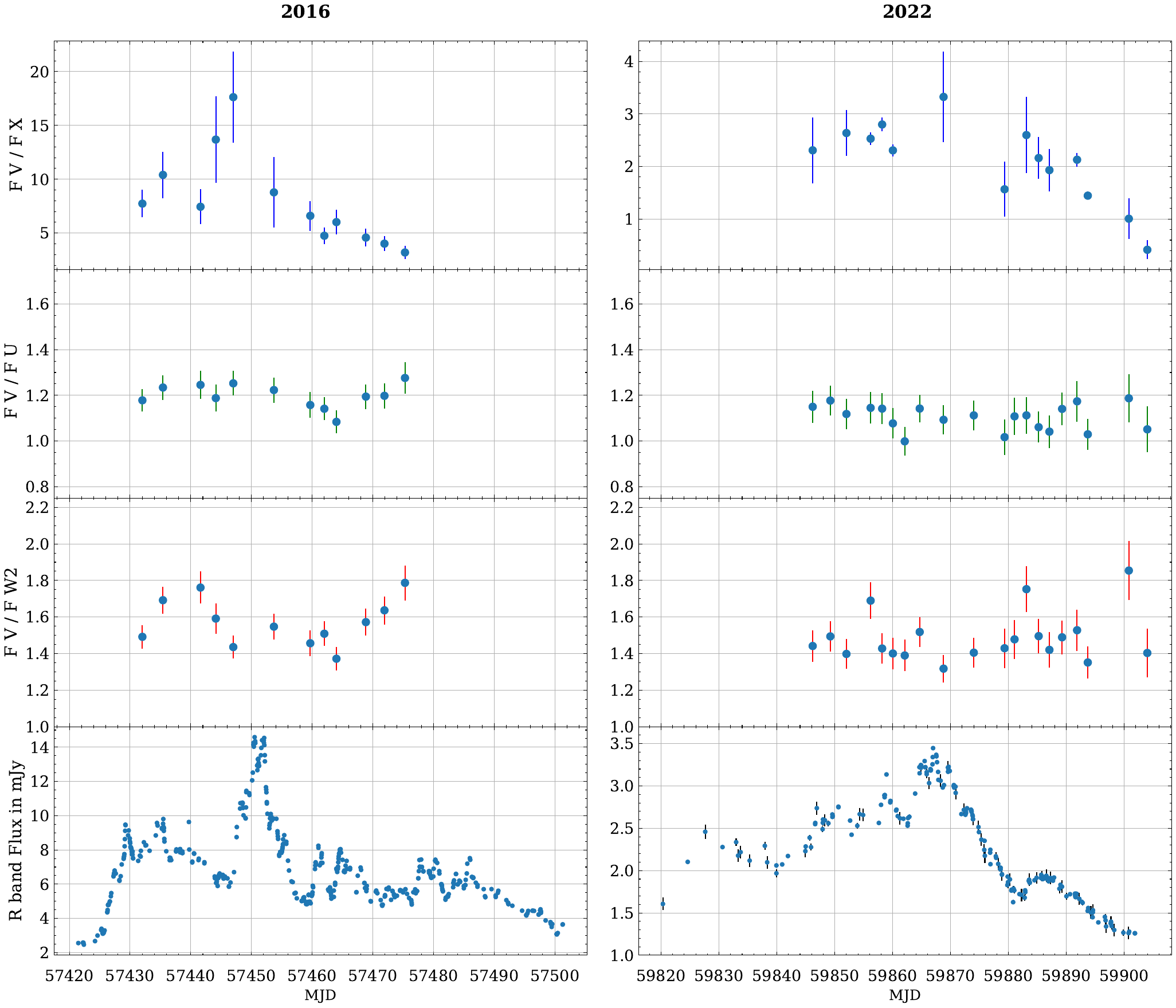} 
    \caption{Swift and R-band monitoring data compared in 2016 and 2022. Theoretically, the 2016 and 2022 flares correspond to each other at two successive apocenter impacts. The top panels gives the V/X light curve for both flares, while the flare light curves are shown by the bottom panels. In both cases there is a significant V/X flare associated with the optical flare. On the other hand, no significant evolution is shown in the V/U and V/W2 flux ratios.} 
    \label{fig:1_2}
\end{figure*}

\begin{figure*}
 \centering
 \includegraphics[scale=0.4]
{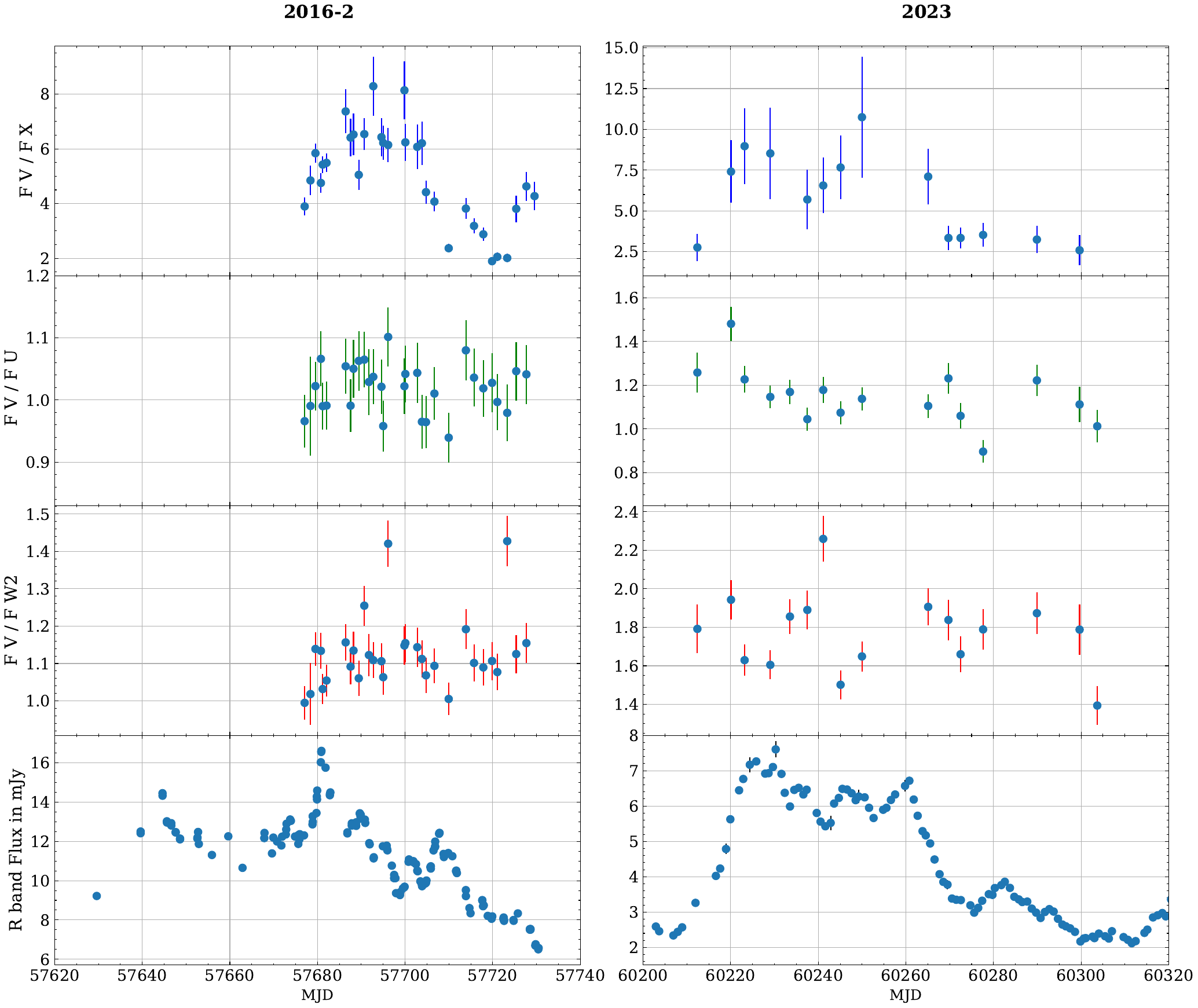} 
    \caption{Swift and R-band monitoring data compared in 2016-2 and 2023. Theoretically, these two flares correspond to each other at two successive apocenter impacts. As in Figure 5, we see strong V/X flares (top panels) associated with the optical flares (bottom panels), while the other flux ratios V/U and V/W2 (middle panels) appear less interesting.} 
    \label{fig:1_4}
\end{figure*}

\begin{figure*}
\hspace{-1cm}
 \centering
 \includegraphics[scale=0.4]
{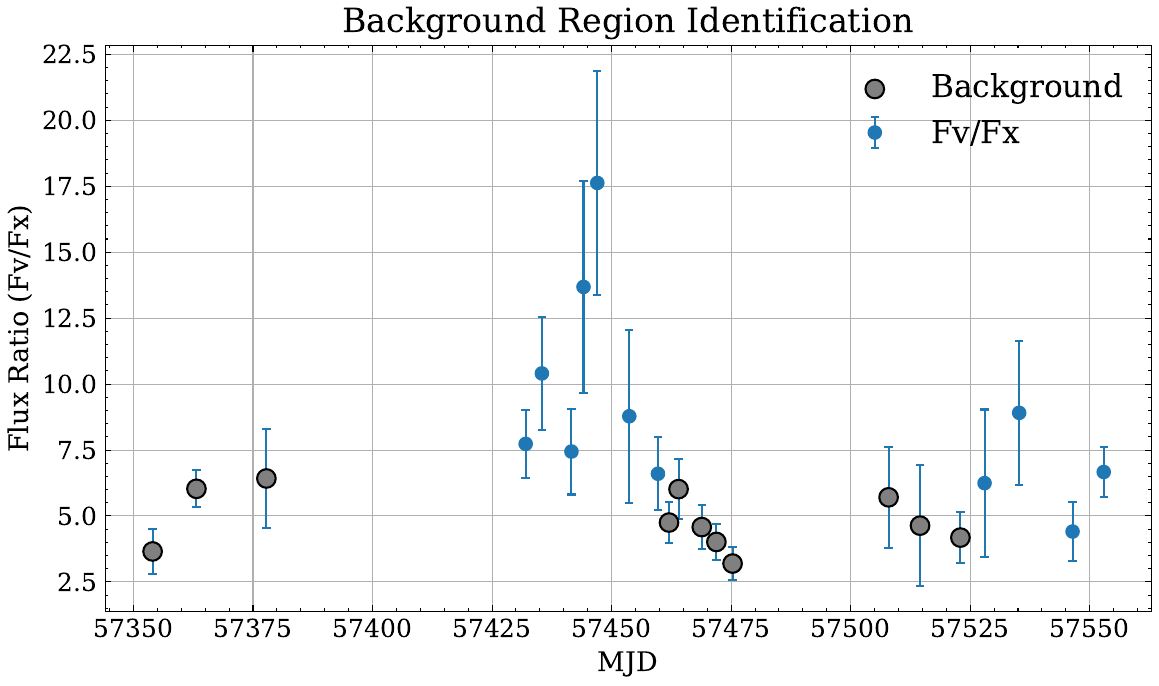} 
   \caption{Swift V/X ratio in 2016, the flare (points) and the background (filled circles).} 
\end{figure*}

\subsection{Photometric, polarimetric and spectral index observations of OJ 287}

OJ~287 is one of the targets included in the Krakow quasar monitoring program. The photometric
measurements of this blazar have been taken at several sites, spanning the entire globe.
Both manually controlled telescopes and robotic ones take part in the program.
The details about the observatories, involved in observations, were given in \cite{2023MNRAS.521.6143V,2024ApJ...968L..17V},
and references therein.
We measure OJ~287 flux in the wide R-band filter regularly leading to a dense light
curve of the target. Multifilter observations are also performed but they are much  
more scarce. Observations of OJ~287 have been done in the following way: we take a series
of measurements in each filter during a night. Subsequently, a mean point is calculated
out of these single points. As a result of the long term monitoring, a differential
light curve is calculated and all sites data use the same comparison star, against which
the target is measured. We adopted star \#4 from \cite{1996A&AS..116..403F} as the comparison
star and \#10 as the check one. We also store all single points and
a mean light curve of OJ~287 can be recalculated with any requested bin width.
In order to investigate colors of the blazar, we report multicolor observations in BVRI
wide band filters, taken during the epochs analyzed in this work. Additionally, we also
consider multifilter data performed in the Sloan photometric system through the griz filters.

University Observatory Jena data were obtained in the R-band with the CCD-imager CTK-II 
\citep{2016AN....337..226M}, using a detector integration time of 180 seconds. All data 
were dark-corrected with dark frames, taken on each observation night, and flat-fielded 
with skyflats, taken during evening or morning twilight. R-band optical data was also 
obtained from The ATLAS Project (Asteroid Terrestrial-impact Last Alert System), as 
described in \cite{2018PASP..130f4505T}.

\begin{figure*}
 \centering
 \includegraphics[scale=0.4,angle=0]
{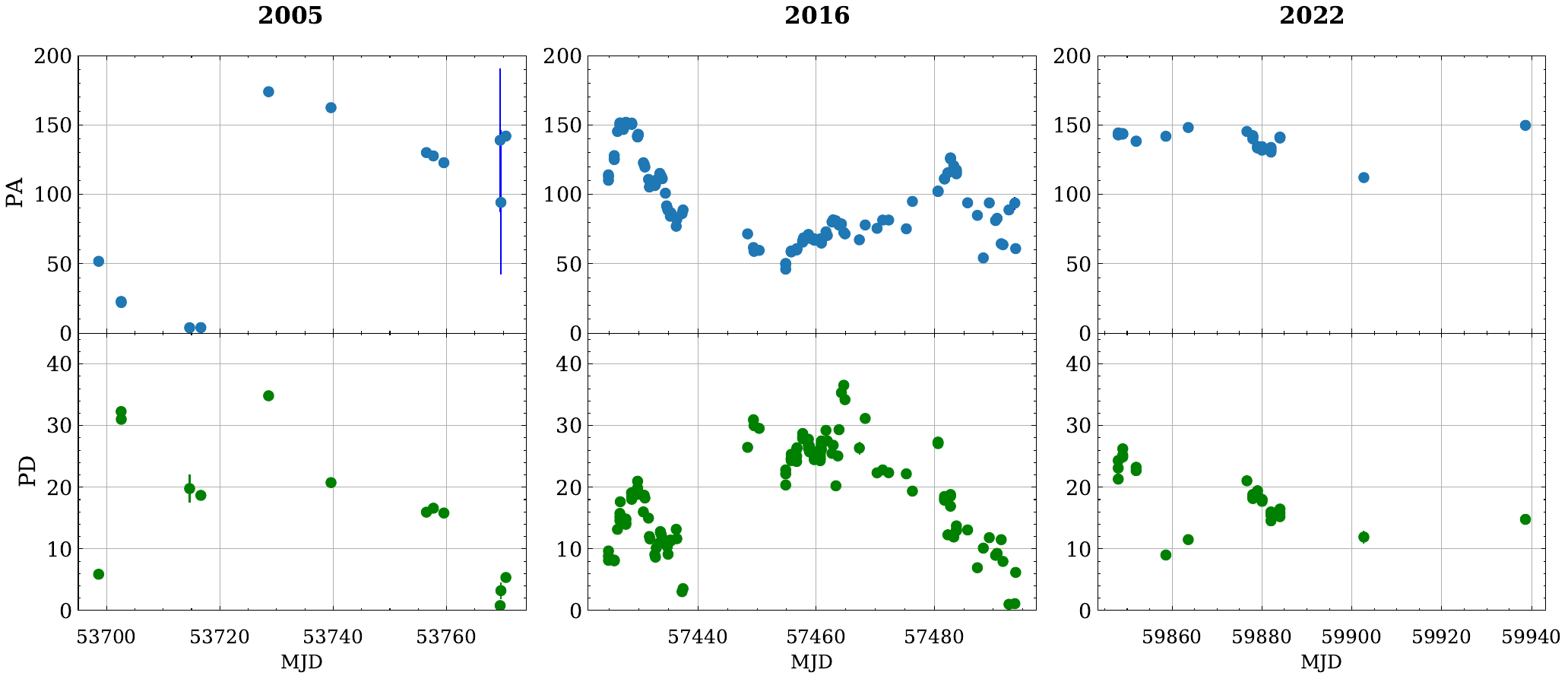} 
    \caption{Polarization in 2005, 2016 and 2022 compared. These three flares correspond to each other and arise theoretically at three successive apocenter impacts. The 2005 and 2022 polarization curves are sparsely sampled, but in any case all three show large degrees of polarization (PD) at least at some phases of flare evolution. Also large swings of the polarization angle (PA) are detected. The 2005 and 2022 flare light curves are displayed in Figure 9.} 
   \label{fig 1_5}
\end{figure*}

\begin{figure*}
\hspace{-1cm}
 \centering
 \includegraphics[scale=0.4,angle=0]
{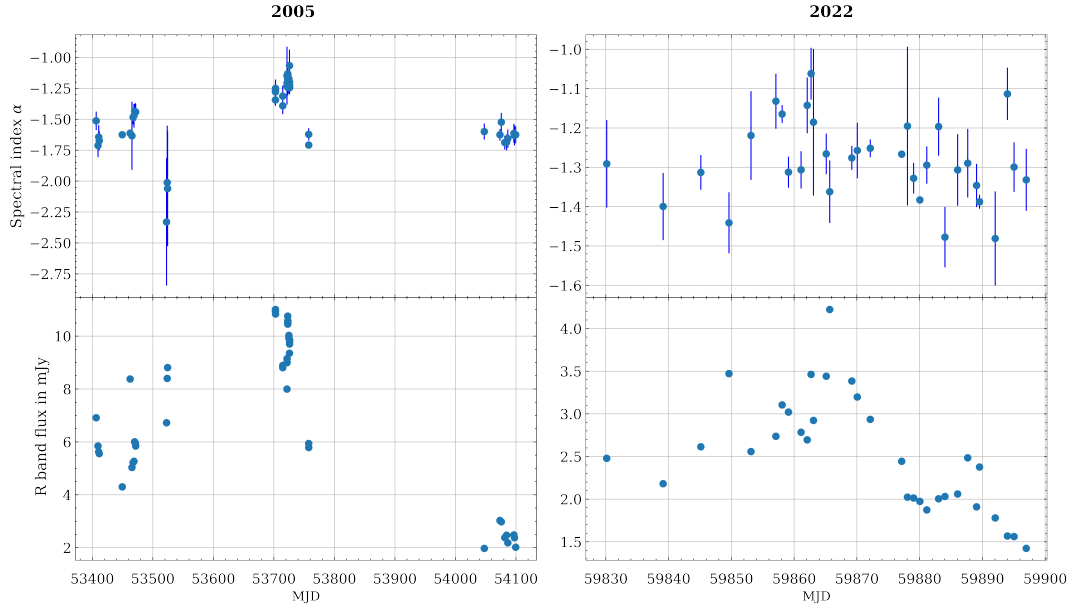} 
    \caption{Spectral index $\alpha$ (top panels) and R-band flux (bottom panels) in 2005 and 2022. These two flares arise at apocenter impacts and should correspond to each other. Both flares show spectral flattening (small value of $-\alpha$) near the peak flux. }
    \label{fig 1_3}
\end{figure*}

The tidal flares, related to disc impacts in the apocentre part of the orbit, are recognized by the double nature of flaring, separated by about one year, by spectral index around -1.2, degree of polarization around 30\%, and a high V/X ratio in Swift observations \citep{2008bves.confE..30C,2010MNRAS.402.2087V,2023ApJ...957L..11G,2023MNRAS.521.6143V}. The 2022 flare agrees with all these signatures.

As we see in Figure 8, the degree of polarization (PD) rose to about 35 degrees in 2005 and 2016 while the position angle of polarization (PA) displayed large swings. In contrast, in 2022 the PD went up to about 25 degrees, also well above the typical 10 percent level. Since the PD is a convolution of the background PD and the flare PD, it is understandable that the observed rise in PD remains smaller when the relative rise in the total flux is less. This the case for the 2022 flare, in comparison with the 2005 and 2016 flares. In all three cases, the intrinsic PD of the pure flare component should be around 40 percent.

In the spectral index comparison of Figure 9 we expect to see the same phenomenon: the flat spectrum ($\alpha \sim-1.0$) flare component causes a greater shift towards a flat overall energy spectrum from the usual $\alpha \sim1.4$ in 2005 than in 2022. Both cases are consistent with the flat intrinsic flare spectral index of $\alpha \sim-1.0$, even though the large scatter in 2022 makes it more difficult to verify.

Also the relative R-band brightness of the 2005 and 2022 flares (the lower panels in Figure 9) is very much as expected from numerical simulations, as we pointed out earlier. Therefore the 2022 October flare belongs most likely to the class of tidal flares.

\section{Hydrogen line correlation}
\label{subsec:hydrogen_line}
\cite{2006ApJ...641..689V}
derived a correlation between the black hole mass $M_{BH}$ and the hydrogen line luminosity $L(H\beta)$ (in units of $10^{42} erg s^{-1}$) and the line width at half maximum $F(H\beta)$ (in units of $1000~ km s^{-1}$) in quasars:

\begin{equation}
M_{BH} = 10^{6.67} [L({H\beta)}]^{0.63} [F(H\beta)]^2 M_{\odot}.                 
\end{equation}

\begin{figure}
\hspace{-1cm}
 \centering
 \includegraphics[scale=0.4,angle=0]
{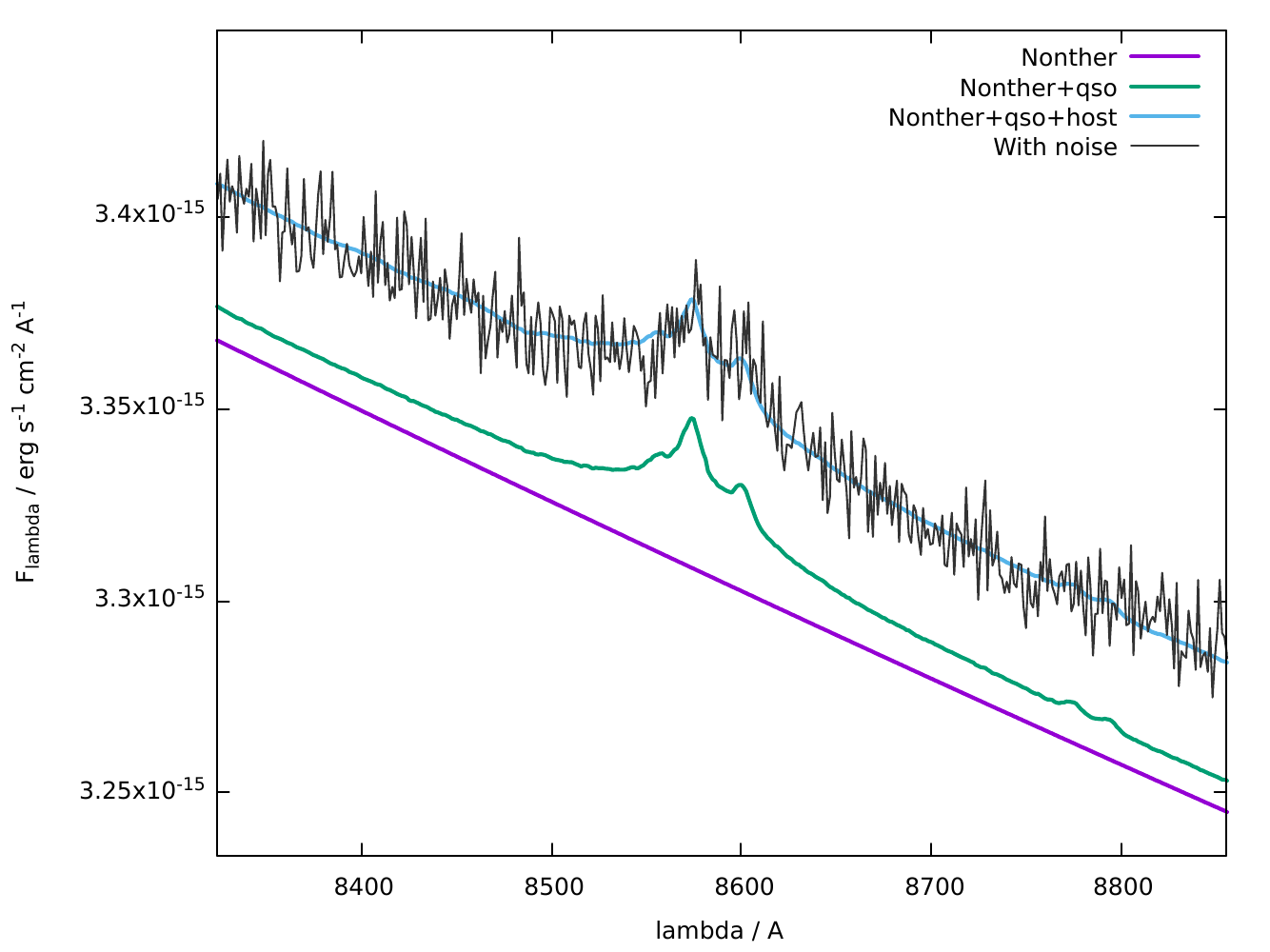} 
    \caption{Decomposition of the April 2008 spectrum of OJ287 near the H$\alpha$ line.} 
    \label{fig 1_6}
\end{figure}

The correlation has scatter, which extends up to a factor six on either side of the mean line. Presumably the origin of the correlation is in the Doppler broadening of the broad line emission region, which is disc-like and viewed at some angle which makes the Doppler shift different from the two sides of the disc, resulting in spectral lines which are broader than the lines from stationary clouds.

Applying this to OJ~287, we need to derive $L(H\beta)$ and $F(H\beta)$ from observations. These quantities are difficult to observe, as the base level of radiation comes from synchrotron radiation of the jet, not from the thermal disc. The hydrogen line has been detected on three occasions when the continuum background level was low. First in 1984 December 23, \cite{1985PASP...97.1158S}
measured  $L(H\alpha) = 10^{42.8} erg s^{-1}$ and $F(H\alpha)$ = $4200 km s^{-1}$  when the V-band flux of OJ~287 was estimated as 0.9 mJy. In 2006 December 6 and 2008 April 8 \cite{2010A&Ap...516A..60N}
measured the value $L(H\alpha)$ on two occasions when the V-band flux was from 1.9 to 4 mJy \citep{2024ApJ...968L..17V}. On five occasions, when the V-band flux was in the range 4 - 8 mJy, the broad $H\alpha$ line was not detected. Similarly, the monitoring from 2010 February to 2013 March, when the brightness OJ~287 was between 2 and 5.5 mJy in V, did not reveal a clear signal of the hydrogen $H\beta$ line \citep{2021Galax..10....1V},
which is expected to be weaker that $H\alpha$ by a factor of three. At its lowest level that OJ~287 has ever been observed, at 0.4 mJy, there is some evidence for the thermal background which, however, may be partly or completely due to the host galaxy in OJ~287 \citep{2022MNRAS.514.3017V}.

In Figure 10 we present a possible break-down of the optical emission contributions from the accretion disc of a $\sim 1.8 \times10^{10}M_{\odot}$ black hole \citep{2023MNRAS.525.1153V}, the host galaxy contribution \citep{2020ApJ...904..102N} and the non-thermal background \cite{2010A&Ap...516A..60N}. The thermal contribution from the disc follows from the work of \cite{2001AJ....122..549V} who
present the average spectral energy distribution for quasars. Since we do not know much about the thermal component in OJ~287, we may assume that its hydrogen $H{\alpha}$ line should appear as in an average quasar.

Note that the hydrogen line strengths did not change between 1984 and 2008. The apparent change 
reported in \cite{2010A&Ap...516A..60N} is due to a misreading of the vertical scale 
in \cite{1985PASP...97.1158S}. K.N. takes this opportunity to acknowledge this error. 

In Eq. 1 we actually need $H\beta$, which was poorly detected in 1984, but dividing $H\alpha$ line flux by 3.06 \citep{2008MNRAS.383..581D}
we get a fair estimate. 
In addition, we should consider how the line-broadening is influenced by the inclination of the accretion disc relative to the plane of the sky. 

The inclination of the accretion disc with respect to the plane of the sky $i$ is likely to be low in OJ~287, say $i \sim 5^{\circ}$, judging from the jet that is coming more or less straight at us.

The definite values of the inclination of the accretion disc are found by \cite{LV96,tateyama04,2005AJ....130.1418J,2009AA...494..527H,VP13,2021MNRAS.503.4400D}: $i = 4.0^{\circ}$, $i = 4.0^{\circ}$, $i = 3.2^{\circ}\pm0.9^{\circ}$, $i = 3.3^{\circ}$, $i = 4.0^{\circ}$, $i = 4.7^{\circ}\pm0.3^{\circ}$, respectively. Many different methods and data sets have been used in different frequencies. We could say that $i = 4.0^{\circ}\pm0.5^{\circ}$ covers the data well.

The Doppler broadening should be proportional to sin$i$. Compared  with a typical value of (sin $i)^{-1}$ in a randomly oriented sample, this correction factor for OJ~287 should be \textbf{$12.4\pm1.5$}, and the increase of the estimated mass by a factor of \textbf{$1.5\pm0.4\times10^2$} over the correlation value. 

Applying Eq. 1 with the inclination correction factor, we get the black hole mass $M_{BH} = (1.9\pm0.5)\times 10^{10} M_{\odot}$. It broadly agrees with the mass determinations based on the infrared-to-UV \citep{2018MNRAS.473.1145K}, radio  \citep{2025ApJ...992...60V} and x-ray \citep{2023A&A...671A.159T} observations.

Figure 10 demonstrates that line emission can be clearly distinguished from jet emission in the Very Large Telescope observations at ESO when the OJ287 is at a low brightness level \citep{2010A&Ap...516A..60N}. Thus,  jet contamination is not a serious issue in this case, although in many other instances it may have been.

We may also note that theoretical stability considerations of the OJ~287 system require the primary black hole mass $M_{BH} \sim 10^{10} M_{\odot}$ \citep{2026ApJ...998..322C}.

In all, the primary mass in OJ~287 determined via the black hole mass - $H\beta$ correlations should be in the range $M_{BH} \sim 10^{10} M_{\odot}$ if OJ~287 is a typical quasar, observed from an atypical viewing angle that makes it a blazar.

\section{Summary and Discussions}
\label{sec:discussion}
We have demonstrated in this paper that a major outburst in OJ~287 occurred in October 2022, and that it corresponds in every way to what was expected of a tidal flare. 

\cite{Komossa_2023b} presented objections concerning this flare. However, they were based on the assumption that it should be an impact flare. On the basis of the data presented here, this was obviously not the case, since in impact flares the degree of polarization decreases, not increases  \citep{val08,val16}. This piece of information was not available to \cite{Komossa_2023b}.

\cite{Komossa_2023b} cite an early draft of \cite{2023MNRAS.521.6143V}, a paper which addresses the question whether it is possible to determine the impact flare arrival epoch without actually observing the impact flare itself, by studying the pre-flare light curve. The 2005 pre-flare lightcurve has a peculiar, up to that time never-before-seen type of flare. It was shown that the time of this flare coincides with the moment when hot, impact shocked gas first emerges from the accretion disc. The question was whether the corresponding feature is also found in the 2022 pre-flare light curve.

Two flares in the 2022 pre-flare light curve were considered, one at JD 2459638 and the other at JD 2459675. For several reasons, the first one was identified as the flare that corresponds to the peculiar 2005 flare. One reason was that the arrival of the 2022 impact flare should be shifted to around 2022 October 10, if the second identification were correct. In the next draft, not distributed widely, it was pointed out that such a three month shift was hardly possible in the standard model with the timing uncertainty of $\pm 1$ week \citep{2024RNAAS...8..276V}.

Unaware of the conclusions of the next draft, \cite{Komossa_2023b} saw a problem in the standard model when an impact flare did not arrive on October 10. They also raised a number of legitimate scientific questions about the standard model which we will discuss below.

One scientific question was related to the primary mass determined by the \cite{2006ApJ...641..689V} correlation that we discuss in Section 4. \cite{Komossa_2023b} did not make any allowance for the inclination correction and therefore their value is about a factor of one hundred smaller than the corrected value. It is simply a lower limit, very far from the true value.

Additionally, \cite{Komossa_2023b} argued against the conclusions of \cite{2021MNRAS.504.5575K} that the x-ray emission observed in OJ~287 is in agreement with the standard black hole mass. They also claimed that the optical accretion disc brightness level in the standard model contradicts observations by overpowering the non-thermal jet emission; we see from our Figure 10 based on theoretical and observational considerations \citep{2023MNRAS.525.1153V} that this is not the case. Neither does the strength of the hydrogen line observed with the VLT \citep{2010A&Ap...516A..60N} present any problem, as we see also from Figure 10, contrary to \cite{Komossa_2023b} arguments.

We show that the tidal flares in OJ~287 follow the same sequence, which has been used to determine the binary orbit, its precession, and therefore the mass of the primary. Already \cite{1997ApJ...484..180S} demonstrated that such flares mark the OJ~287 optical light curve from 1900 to 1996 and predicted the new flares up to 2030. The only exception were the tidal flares arising at the apocenter part of the orbit, which required more refined techniques by \cite{pih13}.

In this paper we have made a detailed comparison of \cite{pih13}
calculations, and compared them with observations. The 2005 disc impact had already happened when \cite{pih13}
carried out their simulations, but the 2016 and 2016/2017 flares as well as 2022 October and 2023 flares constituted predictions at the time. The confirmation of these flares has thus confirmed the model and the primary mass $M_{BH} = (18.35\pm0.05) \times 10^9 M_{\odot}$.

In addition, we have used the black hole mass - hydrogen line width correlation, and combined it with the previously observed at VLT line width in OJ~287. Together with the disk inclination from several independent determinations, the primary mass was calculated. It is remarkable that these disc inclination determinations led to a primary mass value which is fully consistent with the dynamical mass.

Let us discuss the error bars here. The error bars come initially from the orbit solution using the impact flares; they are at the level of $\pm0.01 \times 10^9 M_{\odot}$ \citep{dey18}. The orbit solution uses a Newton-Raphson-type convergent method to find the correct timing for each flare in just a few corrective cycles. The method is accurate and mathematically equivalent to an analytical solution of a ninth order equation. At the same time, it finds the orbit parameters including the primary mass \citep{dey18}. The accuracy of the parameters depends almost entirely on the tolerance in the timings of ten flares, which are given as priors. The method has been developed over more than three decades. In the crudest form, the impacts on a fixed plane (accretion disc) are timed, in later models the effect of the disc geometry was included \citep{val07}. Primary spin was added in \cite{2011ApJ...742...22V,val16}, and finally full radiation reaction terms including the tails were used \citep{dey18}. Interestingly, \cite{2023MNRAS.526.2754Z} dropped all the refinements, and went back almost to the crude model of \cite{LV96}, with only the first order radiation reaction theory, and no disc bending and spin effects. The values of the primary mass shifted from 18.25, 18.40, 18.30, 18.35 and to 18.34 $\times 10^9 M_{\odot}$ at these successive steps. In order to cover the varying number of orbit parameters and the tolerance in priors, we may quote the one sigma error limit $\pm0.05 \times 10^9 M_{\odot}$. 

Note that the two latest mass determinations in the above list give almost exactly the same mass value and uncertainty limits, even though one is only a crude first order model, while the other is a sophisticated model including detailed modeling of discs and gravitational waves. The models differ in many ways, but not in mass. This is because the mass is directly dependent on the precession rate of the binary orbit and its accuracy is related to the tolerance of the timing of the impacts, in particular in the apocenter part of the orbit. In Figure 2 we see that the 2005 impact flare, the highest peak outlined by the dashed line, is easily timed with the 0.02 yr (about one week) accuracy, and this leads directly to the mass error quoted above. The next apocenter flare, shown as the first flare in Figure 3, is similarly sharp and allows the determination of the primary spin value \citep{val16}. The last apocenter flare is placed exactly in the first observational gap on the left hand side in Figure 4. The observational boundaries are so tight that its timing is also determined with the same accuracy as the two previous apocenter flares \citep{2023MNRAS.521.6143V}.

We then come to the question of how well the impacts are timed from tidal flares, in comparison with timing from the impact flares? It is the high precision of the times of the impact flares that gives us the tight error limits for the black hole mass. For the pericenter flares the error in timing is $\pm12$ hours and for the apocenter flares $\pm 1$ week. This leads to the accuracy of $\pm0.01 \times 10^9 M_{\odot}$ in the black hole mass.

For the tidal flares we get an indication of the accuracy from the box sizes used in comparing observations and theory. In our Figure 3 it is four weeks and in Figure 4 it is three weeks. In the \cite{2009ApJ...698..781V} modeling of the 2007 pericenter tidal flare it was two days. Thus we could say that the accuracy is expected to be poorer using tidal flares than using impact flares by a factor of about four. So, if we did not see the impact flares and would be entirely dependent on tidal flares for the orbit solution, the uncertainty in $M_{BH} = (18.35\pm0.05) \times 10^9 M_{\odot}$ would probably refer to one standard deviation. This is something that could be tested in the future, but rather far in the future since the set of three apocenter flares is repeated at 60 year intervals. Of course, it is still possible  that we will find more archival data on the previous sets of apocenter flares at 1896-1906-1913 and 1945-1957-1964 \citep{dey18}.

The tidal flares have an important role especially in the interpretation of the historical light curve. Of the seven brightest flares ever detected in OJ~287, four are impact flares, in 1913, 1957, 1959 and 1983, and three are tidal flares, in 1900, 1907 and in 1971 \citep{2013A&Ap...559A..20H,dey18}. Without tidal flares the triple impact set 1896-1906-1913 could not be observationally verified since in the first two impacts only the tidal component is detected, due to gaps in observations. For the 1945-1957-1964 set, the 1945 impact is based only on the detection of the tidal component. The first impact in the historical record in 1886 is also seen only as a tidal flare.

\section*{Acknowledgements}
This work has made use of data from the Asteroid Terrestrial-impact Last Alert System (ATLAS) project. The Asteroid Terrestrial-impact Last Alert System (ATLAS) project is primarily funded to search for near earth asteroids through NASA grants NN12AR55G, 80NSSC18K0284, and 80NSSC18K1575; byproducts of the NEO search include images and catalogs from the survey area. This work was partially funded by Kepler/K2 grant J1944/80NSSC19K0112 and HST GO-15889, and STFC grants ST/T000198/1 and ST/S006109/1. The ATLAS science products have been made possible through the contributions of the University of Hawaii Institute for Astronomy, the Queen’s University Belfast, the Space Telescope Science Institute, the South African Astronomical Observatory, and The Millennium Institute of Astrophysics (MAS), Chile. This work is based on observations obtained with telescopes of the University Observatory Jena, which is operated by the Astrophysical Institute of the Friedrich-Schiller-University Jena. SC acknowledges support by ASI through contract ASI-INFN 2026-5-HH.0 for SSDC. 



\section*{Data Availability}
The data underlying this article will be shared on reasonable request to the corresponding author.

\bibliographystyle{mnras}
\bibliography{reference} 








\bsp	
\label{lastpage}
\end{document}